# 5G-Enabled Pseudonymity for Cooperative Intelligent Transportation System

Nardine Basta[1], Ming Ding[2], Muhammad Ikram[1], Mohamed Ali Kaafar[1]

*Cooperative Intelligent Transportation Systems (C-ITS) enable communications between vehicles, road-side infrastructures, and road-users to improve users' safety and to efficiently manage traffic. Most, if not all, of the intelligent vehicles-to-everything (V2X) applications, often rely on continuous collection and sharing of sensitive information such as detailed location information which raises privacy concerns. In this light, a common approach to concealing the long-term identity of C-ITS vehicles is using multiple temporary identifiers, called pseudonyms. However, the legacy pseudonyms management approach is prone to linking attacks. The introduction of 5G network to V2X offers enhanced location accuracy, better clock synchronisation, improved modular service-based architecture, and enhanced security and privacy preservation controls. Motivated by the above enhancements, we study 5G-enabled pseudonyms for protecting vehicle identity privacy in C-ITS. We highlight the gaps in the current standards of pseudonyms management. We further provide recommendations regarding the pseudonyms management life-cycle.*

## 1 Introduction

Cooperative Intelligent transportation systems, depicted in Figure 1, enable communication and real-time information sharing among vehicles (V2V), vehicles to roadside infrastructure (V2I), vehicles to pedestrians (V2P), and vehicles to network communications which are collectively known as vehicle-to-everything (V2X) communications. Meanwhile, benefited from the deployment of intelligent transportation systems, traditional transportation conditions are enhanced through the applications of diverse technologies to vehicle infrastructures. The cooperative element in C-ITS is expected to significantly improve road safety, traffic efficiency, and comfort of driving.

To enable this cooperative awareness, vehicles broadcast beacon messages that rely on kinematic data like position and velocity. For instance, Cooperative Awareness Messages (CAM) are sent periodically up to several times a second by every Intelligent Transportation System Station (ITS-S). Similarly, the Decentralized Environment Notification Messages (DENM) are frequently disseminated in order to alert road users of a detected and potentially dangerous event. However, both CAM and DENM messages contain sensitive information (e.g., user or vehicle geo-location) relating to the sending ITS-S. Such data can be used by attackers to track the vehicles and potentially build drivers' profiles (i.e., home/work address, living habits, etc.). Due to the high frequency of up to 10Hz and broadcast nature of ITS safety and awareness messages, a major challenge of the C-ITS is privacy protection. Consequently, privacy-preserving mechanisms need to be devised to hide the true or the long-term identity of vehicles.

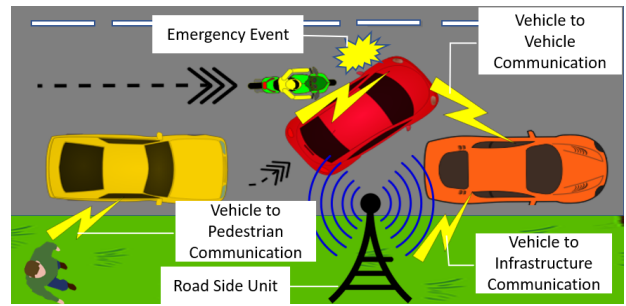

**Figure 1** *Cooperative Intelligent Transportation System (C-ITS) involving vehicles to vehicles communication, vehicles to infrastructure communication and vehicles to pedestrians' communication collectively known as vehicles-to-everything communication (V2X).*

### 1.1 Motivation and Related Work

Concealing the long-term identity of vehicles in C-ITS is crucial for achieving location and identity privacy. However, authentication of messages is still required as forged or manipulated messages could affect V2X-based assistance functions and drivers' safety. Hence, a certain level of link-ability between a vehicle's individual broadcast messages is mandatory for system operation. A common approach to concealing the long-term identity of a vehicle is using multiple temporary identifiers, referred to as pseudonyms. Pseudonyms are public-key certificates that do not include identifying information to prevent location and identity tracking. Despite the availability of alternative approaches, academia, industry and standardization development organizations such as IEEE and ETSI have agreed to adopt the pseudonyms approach to protect the vehicles' location privacy [1].

---

[1] Macquarie University

[2] Data61, CSIRO

The legacy ad-hoc pseudonyms management approaches [1] enable adversaries that eavesdrop on all traffic throughout an area to track the vehicles' trajectories. Eavesdropped messages can be correlated with specific vehicles or even drivers based on reoccurring travel patterns, e.g., commuting trips. The latter is referred to as pseudonyms linking attacks. Two types of linking attacks have been identified: syntactic and semantic linking. The syntactic linking occurs when the attacker tends to break the vehicle privacy through linking its identifiers or pseudonyms by "joining the dots" between two heartbeat messages with the same identifier. The semantic linking, also known as identity revealing attack, is more powerful because it is based on the attacker knowledge of the message content to link the pseudonyms [1].

To mitigate the linking attacks associated with the current pseudonyms management scheme, different approaches have been suggested in the literature, of which the most significant works are: frequent update of pseudonyms, synchronization of pseudonym change [2], the introduction of a radio silence period before the pseudonyms update [2, 3], and the encryption of pseudonyms [4]. Further analysis proved that the silence period approach is more effective than the pseudonyms encryption [1]. Compared to the encryption technique [1], the radio silence provides effective protection against both internal and external passive adversaries, since no safety message is provided to the adversary during the radio silence. On the other hand, with the encryption approach, anyone possessing the credentials can decrypt the intercepted messages. Additionally, equipping vehicles with multiple pseudonyms to enable frequent update of pseudonyms, if concurrent use is not restricted, might allow malicious actors to spoof multiple independent vehicles and conduct a sybil attack. For example, a malicious vehicle can attempt to gain free roads by faking a traffic jam.

While vehicles anonymity is desirable, it conflicts with authorities' wish to ensure accountability and non-repudiation. To address the complex relationship between application requirements, privacy protection, anonymity, authentication, accountability, and non-repudiation requirements, efforts have been made to improve pseudonyms management schemes that aim to balance these divergent requirements [1]. Nevertheless, the drawbacks of the existing approaches raise a valid question on their effectiveness and the level of achievable location privacy protection. The most significant issues are concerning the safety-privacy trade-off. For instance, the work in [2] suggests executing the radio silence-based strategy in places where the impact on road safety is minimal such as regions with widespread road infrastructure [4]. However, many challenges still exist concerning the impact of privacy-preserving measures on the efficiency of disseminating the safety messages, scalability, and communication overhead [1].

*1.2 Problem Statement and Contributions*

Hybrid vehicular communications using different radio technologies and channels are introduced to overcome the shortfalls of ad-hoc communication and increase the channel throughput and reliability. Cellular-V2X (C-V2X) radio technology considers the implementation in different broadband cellular network generations, i.e., LTE-V in 4G and future generations such as 5G. The 5G network in particular, not only is envisioned to enhance the vehicular communication speed and reliability, but also is expected to improve the clock synchronization and the location accuracy (in the order of 1 meter or below) [5].

The 5G system architecture presents an integral change in network architecture. For instance, the introduced Service-Based Architecture (SBA) improves the modularity of the network system where the control plane functionality and common data repositories of a 5G network are represented by a set of interconnected Network Functions (NFs), each with authorization to access each other's services. The SBA allows any third-party application, referred to by 3GPP as Application Function (AF), to interact with the NFs in a secured manner. Hence, it paves the way for more advanced vehicular applications and protocols with security and privacy-preserving features.

To limit pseudonyms linking attacks, in current ETSI standards [6], it is required that each V2X application uses its identifiers that cannot be linked to each other. In particular, DENM and CAM originating from the same vehicle should not be linkable. Accordingly, different pseudonyms should be assigned to the different applications, even associated with the same ITS-S.

ETSI security concept uses long-term certificates for identification and accountability of ITS-S, named Enrolment Certificates (EC) and short-lived, anonymized certificates for V2V/V2I communications, named Authorization Tickets (AT) [6]. A User Equipment (UE) or an ITS-S requires authorization tickets for accessing 5G network services. These authorization tickets have a time validity and are frequently changed. Accordingly, the current ETSI pseudonym management approach relies on the authorization tickets to pseudonymize the ITS-S identity while proving it is authenticated and authorized to access communication resources and services. Efforts have been made by researchers and standardization organizations such as ETSI [6] and SAE [7] to define pseudonyms life-cycle parameters, pseudonyms metrics and update strategies. However, many unresolved issues and open questions still exist.

The contribution of this work is twofold: Firstly, we provide a detailed description of the 5G basic architecture while joining some dots concerning its application in the ITS domain. We present, for the first time, a detailed end-to-end technical overview of the V2X 5G-enabled approach to achieving pseudonimity starting from the vehicle's initial registration with the network phase. We further discuss the 5G security procedures with a focus on authentication and services authorization. Secondly, we discuss the 5G-related unresolved pseudonyms management challenges and suggest recommendations and possible solutions to address these challenges.

## 2 Background

The 5G System architecture brings transformational changes to the core network by improving the modularity of the network system. One key advancement is upgrading the traditional network architecture to Service-Based Architecture (SBA), as shown in Figure 2. Each NF provides a set of services, through a service-based interface that is consumed by other authorized NFs.

An NF service is one type of capability exposed by an NF (NF Service Producer) to other authorized NF (NF Service Consumer) through a service-based interface. The interaction between the consumer and producer takes place within the Service Framework (SF). The SF consists of service registration, service authorization and service discovery. A Service Producer needs to register itself to be visible to consumers. The service authorization ensures the consumer is authorized to access the service provided by the service provider. Finally, the service discovery enables a consumer to discover NF instance(s) that provide the required service(s).

The Network Repository Function (NRF) is the NF of the core network that provides a single record of all network functions (NF) available in a given mobile network, together with their profiles and the services they support. It receives an NF's registration and discovery requests, maintains profiles, and acts as an authorization server where NF Service Consumers can request an OAuth2 [8] token to access the services offered by the designated NF Service Producers. The authorization tokens can be used as a means of achieving pseudonymity as discussed earlier in Section 1.2.

### 2.1 5G UE Enrollment and Primary Authentication

An ITS-S needs to register with the network to get authorized to receive services. The Enrolment (registration) process is the main access control to the ITS. The ITS-S requests its enrolment certificate from the Enrolment Authority (EA). The enrolled ITS-S receives an Enrollment Certificate (EC) which grants it permission to gain authorization for the use of further services. During enrollment, the network performs a primary authentication procedure and provide the keying material to be used between the ITS-S and the serving network in subsequent security procedures. During the primary authentication and key agreement procedure, the serving network authenticates the UE permanent cellular identity, namely the UE Subscription Permanent Identifier (SUPI), which is tied to the subscriber's (U)SIM card [9].

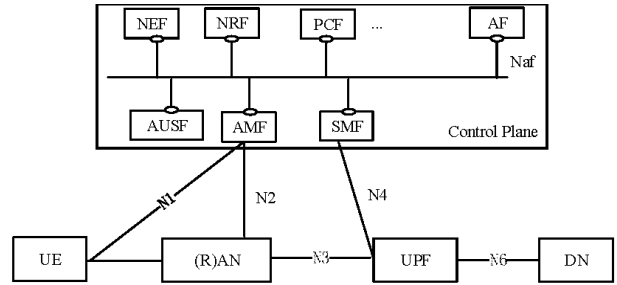

*Figure 2* 5G Service-Based Architecture. The RAN node is the 5G base station. The User Plane Function (UPF) is the gateway node to the Data Network (DN). The control plane functions include the Access and Mobility Management Function (AMF), the Session Management Function (SMF), the Authentication Server Function (AUSF), the Network Repository Function (NRF), the Policy Control Function (PCF), the Application Function (AF) and the Network Exposure Function (NEF).

A major improvement in the 5G authentication procedure is adding the home network's public key to the primary authentication process. With the subscriber UE having the home network's public key stored in the USIM, it can start encrypting sensitive authentication data, e.g., the SUPI, right from the beginning of the authentication process [10].

For an enrolled and authenticated ITS-S to use ITS applications, it should receive cryptographically signed certificates known as the Authorization Tickets (AT) from the Authorization Authority (AA). The ATs guarantee that the requesting ITS-S has the permission to use the requested services. Different variants for the authorization process are defined in the ETSI Standardization documents [6]. The token-based authorization approach using OAuth 2.0 is a well-established way to perform service authorization [9].

### 2.2 5G Service Authorization

The SBA allows any third-party application, referred to by 3GPP as Application Function (AF), to interact with the NFs in a secured manner. The AF represents applications that have been approved by the operators to use the core network either directly or via access to an exposed API. AF consists of two types: trusted (SMS, vehicular applications, voice and video telephony, etc.) and untrusted (Google, Facebook, another roaming operator, etc.). Trusted applications reside in the trusted data network and use the Network Repository Function (NRF) to communicate with the core network. On the other hand, untrusted applications rely on the Network Exposure Function (NEF) for communications with the control plane services to protect the core network.

The 5G system enables Mobile Network Operators (MNO) to (de)authorize UEs and IoT devices to use mobile network services. Accordingly, the 5G system provides means to verify whether a UE is authorized to access a specific service before establishing a direct device connection with the Service Producer NF [11]. The NF Service Producers are various 5G vertical service providers. A UE can subscribe to a variety of services provided by service providers. The UE obtains NF Service Producers' 5G trusted services through the NRF.

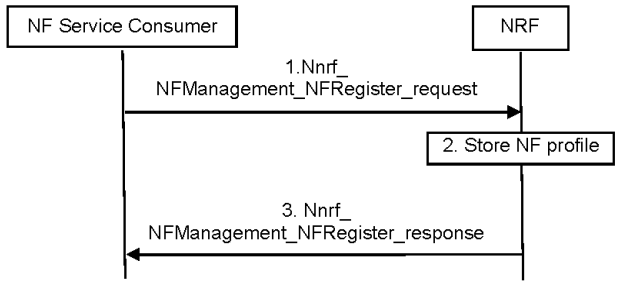

*Figure 3* The registration process between the Network Function (NF) Service Consumer and the Network Repository Function (NRF) [11].

As a 5G service configuration management and authorization server, the NRF can support mutual authentication and service authorization between UEs and NF Service Producers. It is responsible for the registration, discovery and selection of network functions. It further generates OAuth2 authorization tokens for the UEs to enable them to access the requested services.

In the context of C-ITS, a vehicle (UE) requesting V2X service from the V2X Service Producer AF initiates a service request procedure to send uplink signaling messages, user data, or as a response to a network paging request. After receiving the service request message, the Access and Mobility Network Core Function (AMF) plays the role of the Service Consumer NF. For the request to be authorized, the NF Service Consumer (OAuth 2.0 client) registers with the NRF (OAuth authorization server) and requests authorization tokens.

The NRF and NFs authenticate each other during discovery, registration, and access token requests. The authentication can either be by means of Transport Layer Security (TLS) or through client credentials assertion and authentication (JSON Web tokens signed by the NF Service Consumer and secured with digital signatures based on JSON Web Signature (JWS)). The NF-NRF registration process, shown in Figure 3, is described as follows [11]:

1. The NF Service Consumer (AMF) sends *Nnrf_NF_Management_NF* Register Request message to NRF to inform the NRF of its NF profile when the NF Service Consumer becomes operative for the first time.
2. The NRF stores the NF profile of NF Service Consumer and marks the NF Service Consumer available.
3. The NRF acknowledge NF Registration is accepted via *Nnrf_NF_Management_NF* Register Response.

Similarly, the NF Service Producer (V2X AF) registers as OAuth 2.0 resource server in the NRF. The NF profile configuration data of the NF Service Producer may include the "additional scope". The "additional scope" information indicates the resources and the actions (service operations) that are allowed on these resources for the NF Service Consumer. These resources may be per NF type of the NF Service Consumer or per NF instance ID of the consumer [9].

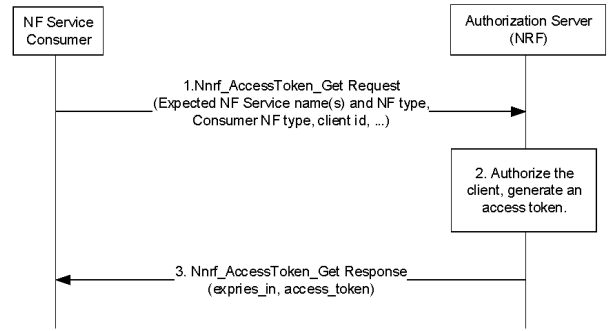

*Figure 4* Access token request process between the Service Consumer Network Function (NF) and the Network Repository Function (NRF). The NRF checks whether the NF is authorised to use the requested service instance and issue authorisation tokens accordingly [9].

After the consumer and producer NFs register and mutually authenticate with the NRF, the consumer is able to send a service request. The complete service request is a two-step process including requesting an access token by NF Service Consumer and then requesting the service based on the granted access token.

### 2.2.1 Access Token Request

The NF Service Consumer obtains an access token before service access to NF Service Producers of a specific NF type. The access token request process is illustrated in Figure 4 and explained as follows [9]:

1. The NF Service Consumer requests access token from the NRF in its same PLMN using the *Nnrf_Access_Token_Get* request command. The request includes the NF Instance Id(s) of the NF Service Consumer, the requested "scope" including the expected NF Service name(s) and optionally "additional scope" information, NF type of the expected NF Service Producer instance and NF Service Consumer.
2. The NRF checks whether the NF Service Consumer is authorized to use the requested NF Service Producer instance/service instance, and then proceeds to generate an OAuth2 access token with the appropriate claims included. The NRF digitally signs the generated access token based on a shared secret or private key as described in [12]. If the NF Service Consumer is not authorized, the NRF does not issue an access token to the NF Service Consumer. Otherwise, the generated token includes the NF Instance Id of NRF (issuer), NF Instance Id of the NF Service Consumer (subject), NF type of the NF Service Producer (audience), expected service name(s), (scope), expiration time (expiration) and optionally "additional scope" information (allowed resources and allowed actions (service operations) on the resources).
3. If the authorization is successful, the NRF sends the access token to the NF Service Consumer in the *Nnrf_AccessToken_Get* response operation. Otherwise, it replies with an OAuth2 error response mentioning the reason for authorization failure [8].

With the butterfly AT provisioning the EA requests authorization tickets on behalf of the ITS-S by communicating with the AA. This approach enables the ITS-S to obtain authorization tickets in batches which reduces the overhead of frequent requests for individual AT.

### 2.2.2 Access Token-Based Service Request

Prior to requesting a service, the consumer NF may perform a network discovery via *Nnrf_NF_Discovery_Request* command to select a suitable NF Service Producer to authorize the service access request. As shown in Figure 5, the NF Service Consumer in possession of a valid access token requests service access from the NF Service Producer following the below steps [9]:
1. The NF Service Consumer requests service from the NF Service Producer while including the access token. The NF Service Consumer and NF Service Producer will then mutually authenticate.
2. The NF Service Producer ensures the token integrity by verifying the signature using NRF's public key or checking the MAC value using the shared secret key. If the integrity check is successful, the NF Service Producer shall verify the claims in the token.
3. If all claims verification is successful, the NF Service Producer grants the requested service and responds back to the NF Service Consumer. Otherwise, it replies with an OAuth2 error response mentioning the reason for authorization failure [8].

For any Service Request, the AMF responds to the UE with a Service Accept message to synchronize PDU Session status between UE and the network, if necessary. The AMF responds with a Service Reject message to UE if the Service Request cannot be accepted by the network. The Service Reject message may include an indication or cause code requesting the UE to perform registration procedure [11].

## 3 V2X Pseudonimity Limitations and Proposals

The SAE and ITSI standardization organizations have provided several recommendations for the pseudonym parameters and change strategies [13, 6]. Nevertheless, many challenges and open questions are yet to be addressed such as the pseudonyms change triggering conditions [6]. In this section we present and discuss some of these challenges, and further provide recommendations for addressing some of the open issues.

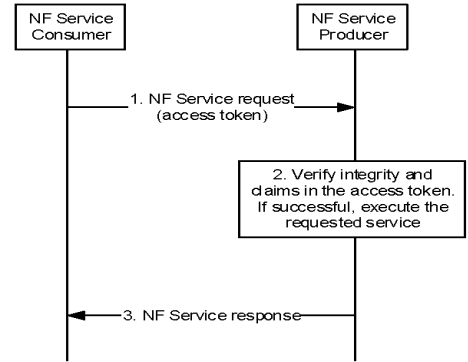

*Figure 5* Service request process between the Network Function Service Consumer and the Network Function Service Producer using the received access tokens [9].

We consider adversaries with no limitations on location tracking capability in any region of interest referred to by [1] as Global Adversaries. We assume adversaries can leverage the ITS infrastructure and RSU in addition to adversarial deployed units. We further consider internal adversaries that are authenticated members of the network as well as external adversaries. Finally, we take into consideration both passive and active adversaries. The former attempts to link multiple pseudonyms to either track vehicle's locations or identify drivers, while the latter tempers with the pseudonyms lifecycles and disturbs the pseudonyms management.

### 3.1 Pseudonyms Update Triggers & Management

Efforts have been spent to provide guidelines and recommendations regarding the prerequisites and timers to trigger the temporary identity re-allocation, which aims to achieve the best trade-off between efficiency and privacy. The SAE for instance recommends updating the pseudonym at startup and then every 5 minutes [6]. The pseudonym change should be accompanied by a silence period to confuse potential eavesdroppers.

The C2C-Communication Consortium approach relies on making location linking infeasible by dividing any road trip into three segments and recommend changing the pseudonym at least once per segment. The first pseudonym change is initiated at the beginning of a trip, the second change should be performed during the trip randomly with 800m – 1500 m from the start of the trip. Subsequent pseudonyms re-allocation is performed at least 800 m from the last pseudonym change and within an additional interval of 2 – 6 minutes [14].

To avoid link-ability, the ETSI and SAE recommend changing all the IDs of the communication stack, including the IPv6 address, synchronously with each pseudonym update. The latter necessitates blocking the emission of messages to avoid revealing some of the stack identities during the update process. Since the pseudonym update process induces additional latency, it is recommended that the latency should be less than 100 *ms* which is the maximum CAM frequency to avoid impacting any ongoing safety applications [6].

While we do not provide numerical studies in this paper, we highlight the following trade-offs and provide suggestions and guidelines to be accounted for during the pseudonyms reallocation process:

- The pseudonyms update should be synchronized using clock synchronization between the vehicles to complicate the syntactic linking attacks. In fact, the impact of synchronization on the enhancing the network resilience to linking attacks highly depends on the synchronization accuracy as well as the number of synchronized vehicles. The improved 5G clock synchronization can be used to improve synchronization accuracy and broaden its scope. GPS synchronization is an alternative, however, it is affected by bad weather and cannot be received in tunnels or near tall buildings [4].
- After concurrently changing their pseudonyms, it is recommended that vehicles go in a silence period and stop sending messages to confuse eavesdroppers and prevent them from tracking the newly acquired pseudonyms. These periods need to be carefully managed to maintain the quality of the network and safety applications. Hence, it is essential to ensure a sufficient number of vehicles that are not in silence period within a certain geographical area to successfully broadcast the safety messages [1].
- To avoid impacting the performance of the ITS applications, a centralized context aware decision of the vehicles within a geographical region that should concurrently change their pseudonyms should be made. Accordingly, we recommend that the call of changing the pseudonym should be made by the visiting network. The base stations should monitor the vehicles within their range and send a signal message to the set of vehicles that are allowed to perform a synchronized update of pseudonyms. We argue that this approach should not incur a high overhead nor complexity as the size of the pseudonym update signal message should be small since it is independent of the pseudonym assignment process described in Section 2.2.1.
- It should be also taken into consideration whether the vehicles are ready to change the pseudonyms, e.g., not downloading a file for vehicle system update and not in a safety critical situation that needs immediate action.
- To enable prioritizing the vehicles when selecting the ones to go in silence period, we recommend labelling the ITS-S with an additional variable to record when the pseudonym was last changed.

## 3.2 Outdated Neighbors' Local Dynamic Map

A vehicle initiating a pseudonym update while broadcasting CAM and DENM messages confuses its neighbors, especially when the pseudonym update is followed by a silence period. These preconditions can result into two possible situations namely ghost vehicles and missing vehicles [6]. In the ghost vehicle case, the neighboring vehicles still have the vehicle initial ID in their Local Dynamic Map (LDM) and assume that the new ID belongs to a new vehicle. Accordingly, they will keep the two records and consider them two different vehicles. While the missing vehicle case is usually caused by the silence period when the vehicles seize messages emission and disappear from the neighbors' LDM. Subsequent to the expiry of the silence period, the vehicle starts sending messages again, hence suddenly becomes visible to its neighbors which may render sudden evasion reactions leading to unsafe situations.

To address the above issues, we suggest that vehicles notify their neighbors to delete the old pseudonym prior to starting the silence period. The difficulty of linking the new pseudonyms to vehicles identity depends on the number of vehicles updating their pseudonyms simultaneously. Clearly, it will have an explicit impact on the standardization because new signaling element of "pseudonym deactivation" needs to be defined for such notifications from the initiator vehicles to its neighbors. Additionally, the impact of these additional messages on the network performance is to be verified.

## 3.3 Pseudonyms Change Locking

In current ETSI road safety application standards [6], it is required to lock the change of pseudonym for a short-term period when a car is in a safety critical situation and might require immediate action. In such situations, initiating a pseudonym update may induce uncertainty and impact road safety, especially if it is followed by a silence period. Hence, the pseudonym lock is introduced through *SN-ID-LOCK* service with a maximum period of 255 seconds [6]. However, the fact that there is no limit for the amount of time an application can renew the lock might render the pseudonym traceable. To mitigate this privacy risk, ETSI recommends differentiating between legitimate applications and 3rd party applications. The former should be authorized to lock the pseudonym change as long as necessary whereas the latter should not. However, in some situations, even legitimate applications can be misled, such as in the case of the ghost neighbor situation motioned earlier, and can be tricked into requesting a lock renewal. As an attempt to limit the infinite pseudonyms locking, the related ETSI standard recommends to specify a maximum *SN-ID-LOCK* time of 15 min in order to prevent infinite pseudonym change locks [6]. It should be noted that the lock should not be renewed beyond the validity time of the pseudonym.

In case of dense networks, when no intermediate action of the vehicle is required, other neighbors can relay the safety messages and the vehicle can safely change its pseudonym. To make such a call, the decision maker must have a global view of the network. Hence, we suggest that a threshold should be adjusted to limit the number or lock request made by a certain application and granted by the security layer. After the threshold is attained, further requests should be validated by the serving network. Regarding the standardization, it means that we need to explicitly set a maximum number of lock requests in the specifications. Further, a mechanism to reset the number of lock requests needs to be defined in the security layer. Again, the impact of this approach on network performance and safety applications should be examined.

### 3.4 Pseudonyms Pool Management

In order to protect the privacy of an ITS-S, a regular change of pseudonym is required. Frequently requesting new pseudonyms to the PKI is inefficient. An alternative approach is having a pool of pre-requested pseudonyms that are available for use for a certain validity period [6]. However, concurrently available pseudonyms can be maliciously used to conduct Sybil attacks. To mitigate the risk of misuse of concurrent pseudonyms, ETSI recommends minimising the number of concurrently valid pseudonym certificates for each ITS-S with a minimum value of 2 in order to allow dynamic timing for pseudonym change.

Further questions that arise in this context are: should pseudonyms reuse be allowed? If yes, what would be the impact on the ITS-S pseudonymity? While we recommend against the reuse of pseudonyms because, if not carefully managed, it might enable vehicles tracking, this decision highly depends on the pool size. Other considerations should be accounted for include the cost of pseudonym change in light of the pseudonym management design and frequency of updating the pseudonyms.

Having a pool of pseudonyms requires further considerations to manage the selection scheme of the next pseudonyms. While project SCOOP [15] proposes a round robin scheme, we suggest that this decision highly depends on whether pseudonyms reuse is allowed in coordination with the frequency of pseudonyms update and the size of the pseudonyms pool.

## 4 Conclusion

Cooperative intelligent transportation systems rely on continuous sharing of vehicles' identities, locations, and other sensitive information. While accountability is an essential ingredient to the safety-critical applications, both identities and locations privacy of vehicles must be preserved. While pseudonyms are proved to satisfy both ITS security and privacy requirements, many issues and challenges of the pseudonyms' management life-cycle are yet to be addressed.

In this work, we discussed ITS security procedures and assessed pseudonyms management issues. To the best of our knowledge, this survey is the first to present an end-to-end detailed overview of the 5G-enabled C-ITS security procedures, to date. To foster further research in this area, we discussed the drawbacks of existing pseudonyms management approaches and complication towards balancing vehicular network privacy-performance trade-off. We further presented recommendations to solve several raised issues.

**References**

[1] Jonathan Petit, Florian Schaub, Michael Feiri, and Frank Kargl. Pseudonym schemes in vehicular networks: A survey. *IEEE Communications Surveys amp Tutorials*, 17:228–255, 03 2015.

[2] Abdelwahab Boualouache, Sidi-Mohammed Senouci, and Samira Moussaoui. Vlpz: The vehicular location privacy zone. *Procedia Computer Science*, 83:369–376, 12 2016.

[3] K. Sampigethaya, L. Huang, M. Li, R. Poovendran, K. Matsuura, K. Sezaki, and WASHINGTON UNIV SEATTLE Dept. of ELECTRICAL ENGINEERING. *CARAVAN: Providing Location Privacy for VANET.* Washington univ Seattle Department of electrical engineering, 2005.

[4] Abdelwahab Boualouache, Sidi-Mohammed Senouci, and Samira Moussaoui. A survey on pseudonym changing strategies for vehicular ad-hoc networks. *IEEE Communications Surveys Tutorials*, 20(1):770–790, 2018.

[5] Ping Zhang, Jian Lu, Yan Wang, and Qiao Wang. Cooperative localization in 5g networks: A survey. *ICT Express*, 3, 03 2017.

[6] Intelligent transport systems (its); security; pre-standardization study on pseudonym change management (etsi tr 103.415 version v1.1.1), 04 2018.

[7] SAE j2735: Dedicated short range communications (dsrc) message set dictionary, 07 2020.

[8] IETF RFC 6749: The oauth 2.0 authorization framework, 10 2012.

[9] 5G; security architecture and procedures for 5g system (3gpp ts 33.501 version 16.3.0 release 16), 08 2021.

[10] David A. Basin, Jannik Dreier, Lucca Hirschi, Sasa Radomirovic, Ralf Sasse, and Vincent Stettler. Formal analysis of 5g authentication. *CoRR*, abs/1806.10360, 2018.

[11] 5g; procedures for the 5g system (5gs) (3gpp ts 23.502 version 15.9.0 release 15), 03 2020.

[12] Michael Jones, John Bradley, and Nat Sakimura. JSON Web Signature (JWS). RFC 7515, May 2015.

[13] SAE j2945: On-board system requirements for v2v safety communications, 04 2020.

[14] C2C-CC. Pki memo v 1.7: C2c-cc public key infrastructure memo. Technical report, 02 2011.

[15] Project-connected vehicles and roads. http://www.scoop.developpement-durable.gouv.fr/en/IMG/pdf/scoop_f_-_presentation_5_april_2018.pdf.

**Nardine Basta** *received M.Sc. degree in informatics from Johannes Kepler University (JKU), Linz, Austria, and Ph.D. degree in informatics from University of Ulm, Germany. She worked on enhancing vehicle's autonomy by designing drivers' behavioral predictive models. She is currently a post-doctoral researcher at Macquarie University Cybersecurity Hub, Sydney, Australia. Her research focuses on IoT and network security and privacy including zero-trust security controls, assessment of enterprise network risks and vehicles' identity privacy issues in 5G enabled ITS.*

**Ming Ding** *(M'12) is a Senior Research Scientist at Data61, CSIRO, Australia. He has authored more than 150 papers in IEEE journals and conferences, and around 20 3GPP standardization contributions. Currently, he holds 21 US patents and has co-invented another 100+ patents on 4G/5G technologies. His research interests include information technology, data privacy and security, machine learning and AI, etc. Currently, he is an editor of IEEE Transactions on Wireless Communications and IEEE Communications Surveys and Tutorials.*

**Muhammad Ikram** *is a lecturer at the School of Computing, Macquarie University. He received the PhD degree from the UNSW. He was a joint Post-Doctoral Research Fellow at the University of Michigan, U.S.A and Macquarie University. He was a visiting scientist at CSIRO.. His current research interests include large-scale measurements, analytics, and analysing security and privacy issues in Web and mobile platforms. He has several publications in prestigious conferences e.g., USENIX, NDSS, PETS, and IMC.*

**Mohamed Ali Kaafar** *is the Executive Director of the Macquarie University Cyber Security Hub, Sydney, Australia. He is a Full Professor at the School of Computing, Faculty of Science and Engineering, Macquarie University. He leads the research and development activities in Cyber Security, Privacy preserving technologies, network measurements and ML security with a focus on data driven approaches. He holds the position of visiting professor of the Chinese Academy of Science (CAS).*